\documentclass[]{tMOP2e}

\usepackage{amsmath,graphicx}

\newcommand{\ket}[1]{| #1 \rangle}
\newcommand{\bket}[1]{\bigl| #1 \bigr\rangle}
\newcommand{\bra}[1]{\langle #1 |}
\newcommand{\bbra}[1]{\bigl\langle #1 \bigr|}
\newcommand{\braket}[2]{\langle #1 | #2 \rangle} 

\makeatletter
\renewcommand{\ps@plain}{%
\renewcommand{\@oddhead}{}%
\renewcommand{\@evenhead}{}%
\renewcommand{\@oddfoot}{\hfil\thepage\hfil}%
\renewcommand{\@evenfoot}{\hfil\thepage\hfil}%
}
\renewcommand{\ps@empty}{%
\renewcommand{\@oddhead}{\small\hfil\thepage}%
\renewcommand{\@evenhead}{\small\thepage\hfil}%
\renewcommand{\@oddfoot}{}%
\renewcommand{\@evenfoot}{}%
}
\renewcommand{\ps@Empty}{%
\renewcommand{\@oddhead}{}%
\renewcommand{\@evenhead}{}%
\renewcommand{\@oddfoot}{\hfil\thepage\hfil}%
\renewcommand{\@evenfoot}{\hfil\thepage\hfil}%
}
\renewcommand{\ps@headings}{%
\renewcommand{\@oddhead}{}%
\renewcommand{\@evenhead}{}%
\renewcommand{\@oddfoot}{\hfil\thepage\hfil}%
\renewcommand{\@evenfoot}{\hfil\thepage\hfil}%
}
\renewcommand{\draftnote}{}
\renewcommand{\@articletype}{\phantom{RESEARCH ARTICLE}}
\makeatother

\begin{document}

\title{How well can you know the edge of a quantum pyramid?%
$^{\ast}$\thanks{$^{\ast}$We dedicate this work to J\'anos Bergou --- 
friend, colleague, and grandmaster of unambiguous discrimination ---
on the occasion of his 60th birthday.}%
}
\author{Berthold-Georg Englert$^{a,b}$ and Jaroslav \v{R}eh\'{a}\v{c}ek$^{c}$
\\\vspace{6pt}  
$^{a}$\textit{Centre for Quantum Technologies, %
National University of Singapore, Singapore 117543, Singapore}\\
$^{b}$\textit{Department of Physics, National University of Singapore, 
Singapore 117542, Singapore}\\
$^{c}$\textit{Department of Optics,
Palacky University, 17.\ listopadu 50, 772\,00 Olomouc, Czech Republic}
\\\vspace{6pt}
\received{04 May 2009}
}

\maketitle

\begin{abstract}
We consider a symmetric quantum communication scenario in which the signal
states are edges of a quantum pyramid of arbitrary dimension and arbitrary
shape, and all edge states are transmitted with the same probability.
The receiver could employ different decoding strategies: he could minimize
the error probability, or discriminate without ambiguity, or extract the
accessible information.
We state the optimal measurement scheme for each strategy.
For large parameter ranges, the standard square-root measurement does not
extract the information optimally.
\end{abstract}

\section{Introduction}
\label{sec:intro}

The problem of discriminating among a given set of quantum states
is an important part of many quantum communication protocols 
\cite{discrim1,discrim2}.
The parameters of quantum communication channels and their
security thresholds with respect to certain eavesdropping attacks are related
to the maximally achievable quality of such discrimination, expressed by the
accessible information associated  with the signal states \cite{accessinfo}. 
Unfortunately, except for rather simple cases, it is not possible to find
explicitly the generalized measurements that maximize the accessed
information.  
The optimality of a given detection scheme is often conjectured and numerical
tools are employed to confirm or reject such a hypothesis.

In this contribution we consider quantum communication between
Alice and Bob, where Alice is using states comprising a quantum
pyramid to encode her message. 
Such a pyramid can be defined by requesting that all pairwise transition
amplitudes between the edge states be the same \cite{pyramid}. 
Bob performs a measurement on the received quantum systems. 
Based on the outcomes he tries to identify the received states and decode the
message sent to him. 
Naturally, Bob seeks the best possible decoding strategy, i.e.
he wants to optimize the measurement on his side so that the accessed
information associated with the signal state is maximal --- in other words: he
wishes to extract the accessible information.
Our choice of quantum pyramids for Alice's signal states
is motivated by the fact that discrimination among 
pyramidal states is regularly encountered in quantum cryptography 
and other quantum information problems, such as Grover's search algorithm
\cite{verifyGrover}, and hence is of considerable practical interest. 
At the same time, the high inherent symmetry of quantum pyramids helps to keep
the calculations relatively simple and permits to arrive at transparent results.

Historically, for some time it was thought that the optimal measurement
strategy for pyramidal states is the well-known square-root 
measurement \cite{srm} and even some security proofs for quantum cryptography
were based on this reasonable conjecture \cite{proofs1,proofs2}.
Recently, however, the optimality of the square-root measurement 
has been invalidated for a range of parameters of acute pyramids 
\cite{englert}, and also for obtuse pyramids with three edges and very small
volume \cite{shor}. 
In this contribution, obtuse pyramids of any dimension are investigated, so
that the optimal measurement strategies for pyramids of any shape and
dimension is established. 
In particular, by exploiting the pyramidal symmetry, and using numerical
methods \cite{rehacek,SOMIM} for confirmation, a family of measurements is
found that outperform the square-root measurement in terms of accessible
information about the edges of obtuse pyramid with few edges and small volume.

The paper is organized as follows. 
We start by introducing quantum pyramids in Sec.~\ref{sec:pyramids}, and then
discuss measurement schemes in general terms in Sec.~\ref{sec:info}.
Section~\ref{sec:SRM} deals with the square-root measurement, which we use as
a benchmark.
Unambiguous discrimination is addressed in Sec.~\ref{sec:UD}, and the
maximization of the accessed information is the theme of Sec.~\ref{sec:IMS},
where we recall the known results about acute pyramids and supplement them
with new insights about obtuse pyramids.
Finally, we present a unified view that regards all occurring measurement
schemes as particular cases of a general scheme and summarize our findings in
a table.

\section{Quantum pyramids}
\label{sec:pyramids}

A quantum pyramid can be defined as a set of $N$ edge states
denoted by normalized kets $|E_j\rangle$, $j=1,2,\dots,N$, such that
all pairwise transition amplitudes among these states are equal and real, 
\begin{equation}
\label{overlaps}
\langle E_j|E_k\rangle=r_0-r_1+N r_1\delta_{jk}\,,
\end{equation} 
where
\begin{equation}
\label{lambda}
 r_0,r_1\ge 0\quad\mbox{and}
 \quad r_0+(N-1)r_1=1\,.
\end{equation}
A simple geometrical picture is obtained by decomposing
edge states in a fixed orthonormal basis so that they can be 
represented by vectors in real $N$-dimensional space;
\begin{figure}
\centerline{\includegraphics[bb=170 520 390 750,clip=]{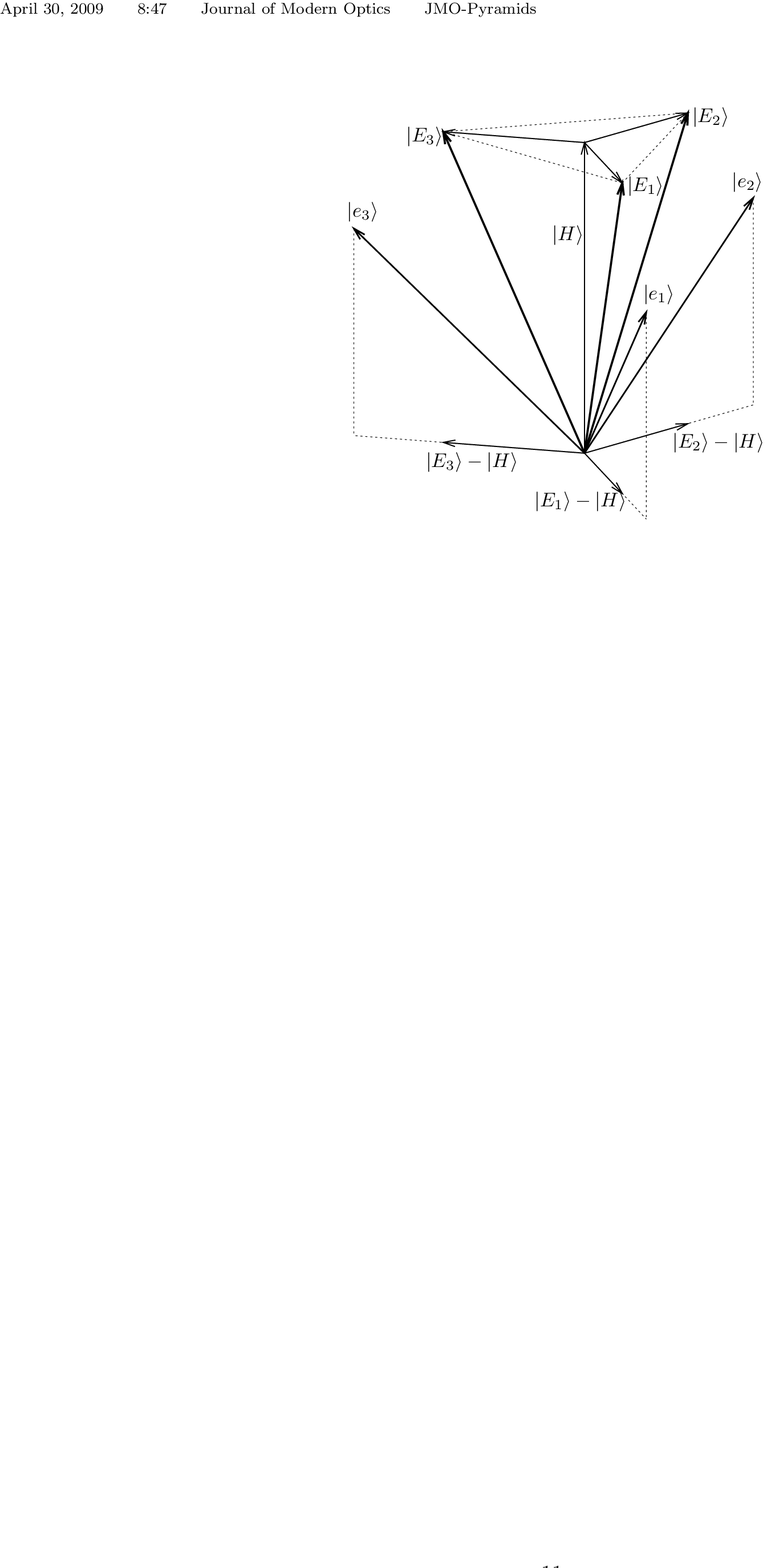}}
\caption{\label{fig:pyramids}% 
Pyramid geometry. 
The three-dimensional cases ($N=3$) is used for illustration.
The pyramid of edge kets $\ket{E_j}$ with $j=1,2,\dots,N$ is depicted as an
object in $N$-dimensional real space. 
The  height ket $\ket{H}$ of (\ref{height}) specifies the symmetry axis of the
pyramid, and the kets ${\ket{E_j}-\ket{H}}$ span the ${(N-1)}$-dimensional
basis of the pyramid.
The orthonormal kets $\ket{e_j}$ of (\ref{orthonormal}) are the edges of the
related orthogonal pyramid, which has maximal volume.
The figure is adopted from Ref.~\cite{englert}.
}
\end{figure}
an example of a quantum pyramid with three edges $|E_1\rangle$, $|E_2\rangle$,
and $|E_3\rangle$ is shown in Fig.~\ref{fig:pyramids}. 

The ``height ket'' 
\begin{equation}
\label{height}
|H\rangle=\frac{1}{N} \sum_j|E_j\rangle\,,\quad 
\langle H|H\rangle=\langle E_j|H\rangle=r_0\,,
\end{equation}
is the pyramidal axis of symmetry, and the kets $\ket{E_j}-\ket{H}$ span the
${(N-1)}$-dimensional base of the pyramid.
Geometrically speaking, the base is itself a ${(N-1)}$-dimensional pyramid
with edges of length $\sqrt{2Nr_1}$ and an angle of $60^\circ$ between the
edges.  

The volume enclosed by a pyramid with $N$ edges, regarded as an object in real
$N$-dimensional Euclidean space, is given by
\begin{equation}
\label{volume}
V=\frac{1}{N!}(N r_0)^{1/2}(N r_1)^{(N-1)/2}\,.
\end{equation}
The single factor $\sqrt{Nr_0}$ refers to the height of the pyramid, and
the ${N-1}$ factors $\sqrt{Nr_1}$ refer to the base.

Pyramids may be classified according to the angle between the edge states.
We call a pyramid \emph{acute} if ${r_0>r_1}$, \emph{orthogonal} if
${r_0=r_1}$, and \emph{obtuse} if ${r_0<r_1}$. 
The orthogonal pyramid ($r_0=r_1=1/N$) has largest volume, and pyramids with
${Nr_0\ll1}$ or ${Nr_1\ll1}$ have very small volume.  
Pyramids with ${r_1 \rightarrow 0}$, are narrow with nearly collinear edges; 
for ${r_1=0}$ we have the degenerate case of an ``all height'' pyramid with no
base. 
The opposite case is the extreme obtuse pyramid with ${r_0= 0}$, 
the ``all base'' pyramid with no height; 
we call such a degenerate pyramid \emph{flat}. 
In Fig.~\ref{fig:pyramids}, the plane pyramid with the three edges
proportional to $|E_j\rangle-|H\rangle$, $j=0,1,2$, 
is an example of a flat pyramid.

Given a pyramid with edges $|E_j\rangle$, the corresponding orthogonal
pyramid with edges $|e_j\rangle$ can be constructed in accordance with 
\begin{equation}
\label{orthonormal}
|e_j\rangle=\bigl(|E_j\rangle-|H\rangle\bigr)\frac{1}{\sqrt{N r_1}}
            +|H\rangle \frac{1}{\sqrt{N r_0}}\,,
\quad\braket{e_j}{e_k}=\delta_{jk}\,,
\end{equation}
which assumes the same orientation with respect to the 
shared axis of symmetry, see Fig.~\ref{fig:pyramids} once more.
An orthogonal pyramid may be \emph{lifted}
by adding a component along the height ket $|H\rangle$,
\begin{equation}
\label{lifted}
|{\bar e}_j\rangle=|e_j\rangle+|H\rangle \frac{t-1}{\sqrt{N r_0}}
\quad\mbox{with}\quad 0\le t < \infty\,.
\end{equation} 
These kets have a genuine $t$ dependence that we leave implicit, 
but the dependence on
$r_0$ is only apparent inasmuch as $\ket{H}/\sqrt{r_0}$ is a unit ket. 
After normalizing them, the lifted edges $|\bar{e}_j\rangle$ make up pyramids 
whose parameters are such that ${\bar{r}_0=t^2\bar{r}_1}$, so that we get
acute pyramids for ${t>1}$, obtuse pyramids for ${t<1}$, and the flat pyramid
for ${t=0}$. 

For future reference, we note that ${t=\sqrt{r_1/r_0}}$ --- for which
${r_0\bar{r}_1=r_1\bar{r}_0}$ --- is particular,
inasmuch as the lifted pyramid for this $t$ value is dual to the original
pyramid, 
\begin{equation}
  \label{dual}
  \braket{E_j}{\bar{e}_k}=\sqrt{Nr_1}\,\delta_{jk}\quad\mbox{for}\quad
  t=\sqrt{\frac{r_1}{r_0}}\,,
\end{equation}
where we need to exclude the degenerate no-volume pyramids with $r_0=0$ or
$r_1=0$. 
Another, more important, observation is the equality
\begin{equation}
  \label{liftsum}
  \sum_j\ket{\bar{e}_j}\bra{\bar{e}_j}=1+\ket{H}\frac{t^2-1}{r_0}\bra{H}\,,
\end{equation}
which is central to what follows because it permits a decomposition of the
identity operator as a sum of nonnegative operators.
We have
\begin{equation}
  \label{1stPOM}
    \sum_j\ket{\bar{e}_j}\bra{\bar{e}_j}+\ket{H}\frac{1-t^2}{r_0}\bra{H}=1
\quad\mbox{for}\quad t\le1
\end{equation}
and
\begin{equation}
  \label{2ndPOM}
    \sum_j\ket{\bar{e}_j}\frac{1}{t^2}\bra{\bar{e}_j}
+\frac{t^2-1}{t^2}\Bigl(1-\ket{H}\frac{1}{r_0}\bra{H}\Bigr)=1
\quad\mbox{for}\quad t\ge1\,,
\end{equation}
which coincide for ${t=1}$.

\section{Accessing information}
\label{sec:info}

\subsection{Measurement schemes}
\label{sec:IMS-MEM-MUD}

The generic scenario of quantum communication is as follows.
Alice sends Bob quantum systems prepared in states described by the normalized
statistical operators $\rho_1,\rho_2,\dots,\rho_j,\dots,\rho_N$, 
with prior probabilities $p_{j\cdot}$ of unit sum.
Bob examines the received systems one by one with the aid of a generalized
measurement with outcomes $\Pi_k$, $k=1,2,\dots,K$; typically ${k\ge N}$.
The outcomes are nonnegative probability operators that decompose the
identity \cite{range},%
\begin{equation}
  \label{genPOM}
  \sum_k\Pi_k=1\,,
\end{equation}
so that the joint probabilities $p_{jk}$ of state $\rho_j$ being
sent \emph{and} outcome $\Pi_k$ being detected have unit sum,
\begin{equation}
  \label{unitsum}
  p_{jk}=p_{j\cdot}\mathrm{Tr}(\rho_j\Pi_k)\ge0\,,\quad\sum_{j,k}p_{jk}=1\,,
\end{equation}
as they should.
The outcomes $\Pi_k$ constitute a \underline{p}robability \underline{o}perator
\underline{m}easurement (POM) in the physics jargon, whereas the acronyom
POVM, short for \underline{p}ositive \underline{o}perator-\underline{v}alued
\underline{m}easure, is preferred in the more mathematically oriented
literature. 
We side with the physicists. 

The information accessed by the POM in use is quantified by 
the mutual information between Alice and Bob which --- in units of bits --- is
given by
\begin{equation}\label{funct}
I=\sum_{j,k} p_{jk} \log_2\frac{p_{jk}}{p_{j\cdot} p_{\cdot k}}\,,
\end{equation}
where 
\begin{equation}
  \label{marginals}
  p_{j\cdot}=\sum_kp_{jk}\,,\quad p_{\cdot k}=\sum_jp_{jk}
\end{equation}
are the corresponding marginal probabilities.
If Bob chooses his POM such that the value of $I$ is maximized, he implements
the \emph{information maximizing scheme} (IMS) and thus extracts the
\emph{accessible} information. 

Quite a few properties of IMSs are known, but there are also many open
questions; a recent review of the matter is Ref.~\cite{IMSreview}.
The calculation of the accessible information involves a multidimensional
nonlinear optimization of the mutual information \eqref{funct} 
with respect to the POM outcomes $\Pi_k$. 
This is a difficult problem and closed-form solutions are known only for
situations in which there is much symmetry among the signal states  
\cite{shor,holevo,sasaki,convexity}; we are adding one item to this list here.
When such solutions are not at hand, one must often rely on a
numerical search, possibly by the iteration method of Ref.~\cite{rehacek} for
which an open-source code is available~\cite{SOMIM}. 
The numerical methods are also useful for the verification or rejection of
conjectured solutions. 

Bob could have other objectives than extracting the accessible information.
For instance, if he wishes to maximize his odds of winning bets on the state
sent by Alice, he implements the \emph{measurement for error minimization}
(MEM), which has as many outcomes as there are states (${K=N}$) and maximizes
$\sum_jp_{jj}$, the probability of guessing the signal state right.
As a necessary condition on the respective POMs we have \cite{holevo}
\begin{equation}
  \label{necessary}
  \Pi_kR_k\Pi_l=\Pi_kR_l\Pi_l\quad\mbox{for all $k,l=1,2,\dots,K$}
\end{equation}
with
\begin{equation}
  \label{defR}
  R_k=\left\{
    \begin{array}{c@{\ \mbox{for the}\ }l}
      \displaystyle
      \sum_jp_{j\cdot}\rho_j\log_2\frac{p_{jk}}{p_{j\cdot} p_{\cdot k}}
      &\mbox{IMS,}\\[1ex]
      p_{k\cdot}\rho_k&\mbox{MEM,}
    \end{array}\right.
\end{equation}
but there is an unfortunate lack of equally powerful sufficient conditions.

Yet another scheme is the \emph{measurement for unambiguous discrimination}
(MUD), which is such that outcomes $\Pi_k$ with $k=1,2,\dots,N$ imply that the
$k$th state was sent, that is: $p_{jk}=p_{\cdot k}\delta_{jk}$ 
for $k=1,2,\dots,N$, whereas the outcomes $\Pi_{N+1},\dots,\Pi_K$ are
inconclusive \cite{unambiguous1,unambiguous2}.
Unambiguous discrimination is not possible for arbitrary signal states
$\rho_j$ and, as a rule, the POMs needed for the IMS, the MEM, and the MUD are
not the same.

\subsection{Symmetric pyramid communication}
\label{sec:pyracommun}

Specifically, we are here interested in symmetric communication with pyramid
states, where $\rho_j=\ket{E_j}\bra{E_j}$ and $p_{j\cdot}=1/N$ for
$j=1,2,\dots,N$. 
Owing to the high symmetry of this situation, the POMs for IMS, MEM, and MUD
can be stated quite explicitly, both for acute and obtuse pyramids. 

The symmetry at hand is, of course, the cyclic nature of the pyramid states:
there is a unitary operator $U$ with period $N$ that cyclically permutes the
pyramid states,
\begin{equation}
  \label{cycle}
  U\ket{E_j}=\ket{E_{j+1}}\quad\mbox{for $j=1,2,\dots,N-1$}\,,
  \qquad U\ket{E_N}=\ket{E_1}\,.
\end{equation}
Explicit expressions for $U$ are
\begin{equation}
  \label{cycleU}
  U=\sum_j\ket{e_{j+1}}\bra{e_j}
   =\sum_j\ket{E_{j+1}}\frac{1}{Nr_1}\bra{E_j}
    +\ket{H}\Bigl(\frac{1}{r_0}-\frac{1}{r_1}\Bigr)\bra{H}\,,
\end{equation}
where $\ket{e_{N+1}}\equiv\ket{e_1}$ and $\ket{E_{N+1}}\equiv\ket{E_1}$ are
understood.  
The ensemble of equal-weight pyramid states is, therefore, an example of the
``group covariant case'' of section 11.4.4 in Ref.~\cite{IMSreview}.

\section{Error minimization: The pretty good square-root measurement}
\label{sec:SRM}

There is a family of POMs that are directly constructed from the signal
states $\rho_j$, their weights $p_{j\cdot}$, and their weighted sum,
\begin{equation}
  \label{rho}
  \rho=\sum_jp_{j\cdot}\rho_j\,,
\end{equation}
which is the statistical operator for the state in which Bob receives the
quantum systems.
For any operator $Z$ such that $Z^{\dagger}\rho Z=1$, the $N$ outcomes
\begin{equation}
  \label{Zroot}
  \Pi_k=Z^{\dagger}p_{k\cdot}\rho_k Z \quad\mbox{for $k=1,2,...,N$}
\end{equation}
make up a POM with $K=N$.
In particular, if $Z$ is chosen nonnegative, that is:
$Z=Z^{\dagger}=\rho^{-1/2}$, we have the so-called 
\emph{square root measurement} (SRM), also known as the ``pretty good
measurement'' because it often accesses a good fraction of the accessible
information.
All other permissible choices for $Z$ are of the form $Z=\rho^{-1/2}U$ with
unitary $U$ and amount to an over-all unitary transformation of the SRM
\cite{isometry}. 

In the situation of present interest, the symmetric communication with pyramid
states, the SRM is the von Neumann measurement composed of the projectors onto
the edges of the orthogonal pyramid \cite{srm}, and the SRM coincides with the
MEM, 
\begin{equation}
  \label{SRM}
  \text{SRM}\equiv\text{MEM}:\qquad 
   \Pi_k=\ket{e_k}\bra{e_k}\quad\mbox{for $k=1,2,\dots,N\,.$}
\end{equation}
The resulting odds for guessing the signal state right are
\begin{equation}
  \label{SRModds}
  \sum_jp_{jj}=\sum_j\frac{1}{N}\bigl|\braket{E_j}{e_j}\bigr|^2
             =\frac{1}{N}\bigl(\sqrt{r_0}+(N-1)\sqrt{r_1}\,\bigr)^2\,,
\end{equation}
which equals $1/N$, $1$, and $1-1/N$ for the no-base pyramid ($r_1=0$), the
orthogonal pyramid ($r_0=r_1=1/N$), and the flat pyramid ($r_0=0$),
respectively. 

For some time, it was conjectured that the SRM is also identical with the IMS
but, as we shall see below, this is only the case for pyramids with
sufficiently large volume.
In view of this historical conjecture, we use the information accessed by the
SRM, 
\begin{equation}
\label{SRMinfo}
\begin{split}
I^{(\text{SRM})}=&\frac{1}{N}\bigl(\sqrt{r_0}+(N-1)\sqrt{r_1}\,\bigr)^2
\log_2\bigl(\sqrt{r_0}+(N-1)\sqrt{r_1}\,\bigr)^2\\
&+\frac{N-1}{N}\bigl(\sqrt{r_0}-\sqrt{r_1}\,\bigr)^2 
  \log_2\bigl(\sqrt{r_0}- \sqrt{r_1}\,\bigr)^2\,,
\end{split}
\end{equation}
as a benchmark value.
For $r_0=r_1=1/N$, we have the maximal value of $I^{(\text{SRM})}=\log_2N$, 
as it should be, and the small-volume limiting values
\begin{equation}
  \label{limitSRMinfo}
  I^{(\text{SRM})}=\left\{
  \begin{array}{c@{\ \mbox{for}\ }l}
    \displaystyle\frac{2N-2}{\ln2}r_1 
    &\displaystyle r_1\ll\frac{1}{N}\,,\\[2ex]
    \displaystyle\frac{\log_2(N-1)}{N}\bigl(N-2+4\sqrt{(N-1)r_0}\,\bigr)
    &\displaystyle r_0\ll\frac{1}{N}
  \end{array}\right.
\end{equation}
are worth noting as well.

\section{Unambiguous discrimination}
\label{sec:UD}

We achieve unambiguous discrimination with the aid of the POMs (\ref{1stPOM})
and (\ref{2ndPOM}) for the particular $t=\sqrt{r_1/r_0}$ value of (\ref{dual}).
These POMs have $N+1$ outcomes, of which the $(N+1)$th is inconclusive
\cite{chefles}, 
\begin{eqnarray}
  \label{MUD}
  \Pi_k&=&\left\{
    \begin{array}{c@{\ \mbox{for}\ }l}
      \displaystyle \ket{\bar{e}_k}\bra{\bar{e}_k} 
      &\displaystyle r_0>\frac{1}{N}>r_1\\[2ex]
      \displaystyle \ket{\bar{e}_k}\frac{r_0}{r_1}\bra{\bar{e}_k} 
      &\displaystyle r_0<\frac{1}{N}<r_1      
    \end{array}\right\}\quad\mbox{for $k=1,2,\dots,N\,,$}
\nonumber\\
\Pi_{N+1}&=&\left\{
    \begin{array}{c@{\ \mbox{for}\ }l}
      \displaystyle \ket{H}\frac{r_0-r_1}{r_0^2}\bra{H} 
      &\displaystyle r_0>\frac{1}{N}>r_1\\[2ex]
      \displaystyle\frac{r_1-r_0}{r_1}\Bigl(1-\ket{H}\frac{1}{r_0}\bra{H}\Bigr) 
      &\displaystyle r_0<\frac{1}{N}<r_1      
    \end{array}\right\}\,.
\end{eqnarray}
The probability that the unambiguous discrimination fails is given by the
probability of getting the inconclusive outcome,
\begin{equation}
  \label{MUDfail}
  \sum_j\frac{1}{N}\bra{E_j}\Pi_{N+1}\ket{E_j}
=\left\{
  \begin{array}{c@{\ \mbox{for}\ }l}
    \displaystyle \frac{Nr_0-1}{N-1} & 
    \displaystyle 1\ge r_0>\frac{1}{N}\mbox{\ (acute),} \\[2ex]
    1-Nr_0 &
    \displaystyle 0\le r_0<\frac{1}{N}\mbox{\ (obtuse),} 
  \end{array}\right.
\end{equation}
and, of course, there is no failure for the orthogonal pyramid with $Nr_0=1$.

The information accessed by the MUDs of (\ref{MUD}) is 
\begin{equation}
  \label{MUDinfo}
  I^{\text{(MUD)}}=\left\{\begin{array}{ll}
  Nr_1\log_2N & \mbox{for acute pyramids,}\\[1ex]
  Nr_0\log_2N & \mbox{for obtuse pyramids.}
 \end{array}\right.
\end{equation}
But this does not do full justice to the obtuse case, where the inconclusive
outcome $\Pi_{N+1}$ is proportional to the projector on the
$(N-1)$-dimensional pyramid base, and there is more information to be accessed
by a decomposition of this $\Pi_{N+1}$ into rank-1 outcomes. 
We introduce the normalized \emph{difference kets}
\begin{equation}
  \label{defdiffket}
  \bket{[mn]}=\bigl(\ket{E_m}-\ket{E_n}\bigr)\frac{1}{\sqrt{2Nr_1}}
\quad\mbox{with $1\leq m<n\leq N$}\,,
\end{equation}
which are $N(N-1)/2$ in number and offer a decomposition of the projector onto
the pyramid base,
\begin{equation}
  \label{sum[mn]}
  1-\ket{H}\frac{1}{r_0}\bra{H}=\frac{2}{N}\sum_{m<n}\bket{[mn]}\bbra{[mn]}\,.
\end{equation}
For obtuse pyramids, then, we can replace the $(N+1)$th outcome of (\ref{MUD})
by $N(N-1)/2$ more informative outcomes in accordance with
\begin{equation}
  \label{MUD'}
  r_1>r_0\,:\quad \Pi_{N+1}=\sum_{m<n}\Pi_{[mn]}\quad\mbox{with}\enskip
\Pi_{[mn]}=\bket{[mn]}\frac{2(r_1-r_0)}{Nr_1}\bbra{[mn]}\,.
\end{equation}
For these we have the joint probabilities
\begin{equation}
  \label{MUD''}
    r_1>r_0\,:\quad 
              p_{j[mn]}=\frac{r_1-r_0}{N}\bigl(\delta_{jm}+\delta_{jn}\bigr)\,,
\end{equation}
so that upon getting the $[mn]$th outcome, Bob knows that there is an equal
chance for Alice having sent the $m$th or $n$th signal state, but it must
surely have been one of the two. 

This makes no difference to unambiguous discrimination because a much reduced
ambiguity --- one of two rather than one of $N$ --- is still an ambiguity, but
it gives $\log_2(N/2)$ bits of accessed information.
\begin{figure}
\centerline{\includegraphics[bb=85 487 475 742,clip=]{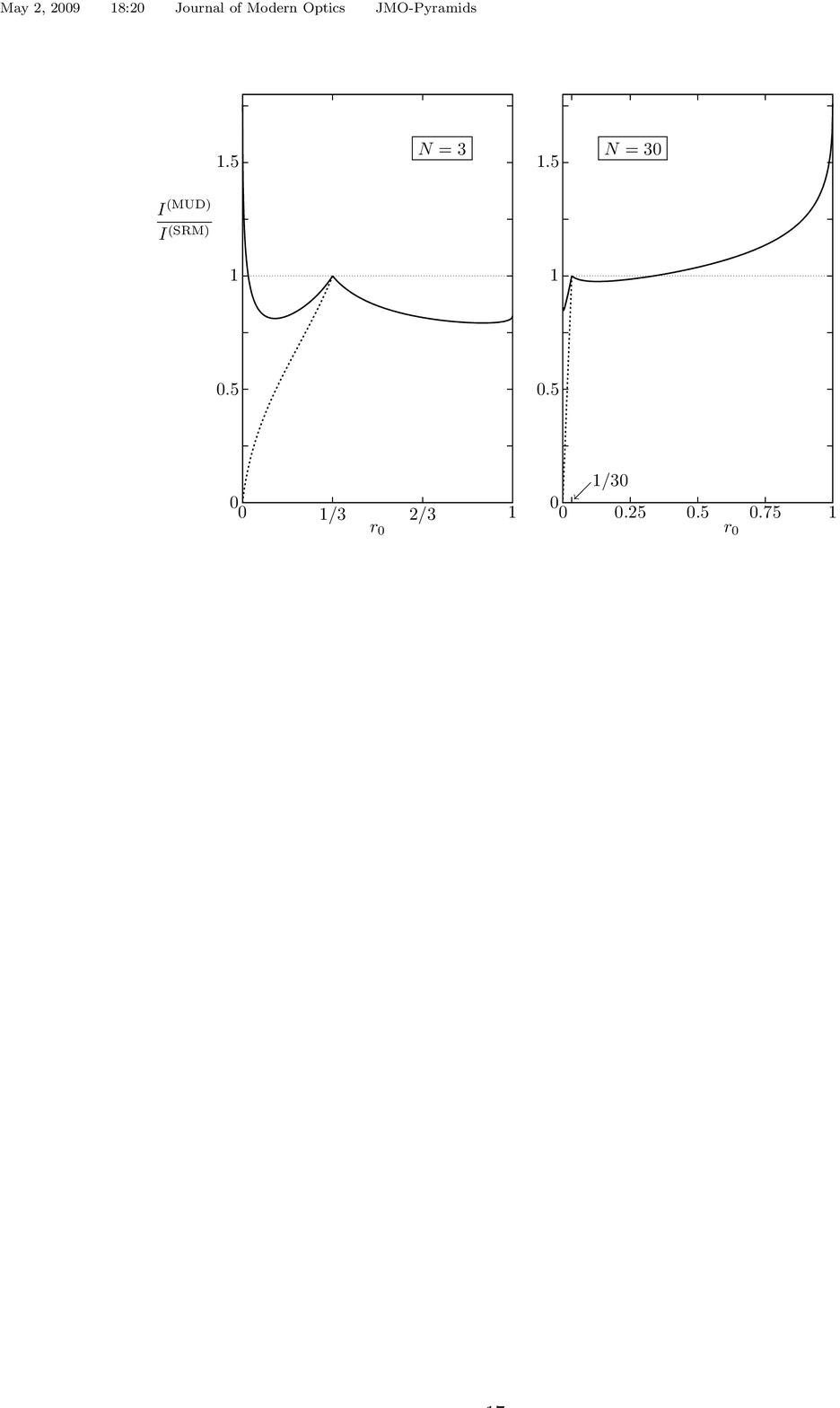}}
\caption{\label{fig:MUDvsSRM}% 
Comparison of the information accessed by the MUD and by the SRM.
For $N=3$ and $N=30$ the solid lines show $I^{\text{(MUD)}}$ of
(\ref{MUDinfo'}) in units of  $I^{\text{(SRM)}}$ of (\ref{SRMinfo}).
For obtuse pyramids, $r_0<1/N$, the dashed lines indicate the value of
(\ref{MUDinfo}) for the less informative MUD.
}
\end{figure}
The resulting improvement over (\ref{MUDinfo}) is
\begin{equation}
  \label{MUDinfo'}
  I^{\text{(MUD)}}=\left\{\begin{array}{cl}
  Nr_1\log_2N & \mbox{for $r_0>r_1\,,$}\\[1ex]
  Nr_0\log_2N+(1-Nr_0)\log_2(N/2) & \mbox{for $r_1>r_0\,,$}
 \end{array}\right.
\end{equation}
which we compare with $I^{\text{(SRM)}}$ in Fig.~\ref{fig:MUDvsSRM}.
As the $N=3$ and $N=30$ examples of the figure show, the MUD may access more
information than the SRM if the pyramid has small volume.
Indeed, the limiting values
\begin{equation}
  \label{MUD/SRMlimit}
  \frac{I^{\text{(MUD)}}}{I^{\text{(SRM)}}}\longrightarrow\left\{
    \begin{array}{c@{\ \mbox{for}\ }l}
    \displaystyle\frac{N}{2N-2}\ln N
    & r_0\to1\,,\\[2ex]
    \displaystyle\frac{N}{N-2}\frac{\ln(N/2)}{\ln(N-1)}
    & r_0\to0  
    \end{array}\right.
\end{equation}
show that $I^{\text{(MUD)}}>I^{\text{(SRM)}}$ for small-volume acute pyramids
with more than four edges and for small-volume obtuse pyramids with less than
seven edges.
It follows that the SRM cannot be the IMS for these pyramids, so that the
historical conjecture mentioned in the context of 
(\ref{SRMinfo}) is definitively rejected for such small-volume pyramids.

\section{Maximizing the accessed information}\label{sec:IMS}
\subsection{Acute pyramids}
\label{sec:acute}

The observation that there are parameter regimes in which the IMS is
not identical with the SRM raises the question of which POM realizes the IMS.
For acute pyramids ($r_0>r_1$), the answer is given in Ref.~\cite{englert}:
optimize the value of $t$ in (\ref{1stPOM}).
This gives
\begin{equation}
  \label{IMSacute}
  \Pi_k=\ket{\bar{e}_k}\bra{\bar{e}_k}\quad\mbox{for $k=1,2,\dots,N$}\,,
\qquad \Pi_{N+1}=\ket{H}\frac{1-t^2}{r_0}\bra{H}
\end{equation}
with
\begin{equation}\label{OPTacute}
  t=\text{min}\left\{1, \frac{2N-2}{N-2}\sqrt{\frac{r_1}{r_0}}\right\}\,,
\end{equation}
so that ${t=1}$ and $\text{IMS}=\text{SRM}$ for the acute pyramids with
${1<Nr_0<4-4/N}$, but not for those with ${Nr_0>4-4/N}$. 
Note that the IMS is always different from the MUD because the $t$ value for
the IMS is always larger than the value $t=\sqrt{r_1/r_0}$ of the MUD.

\begin{figure}
\centerline{\includegraphics[bb=135 565 420 745,clip=]{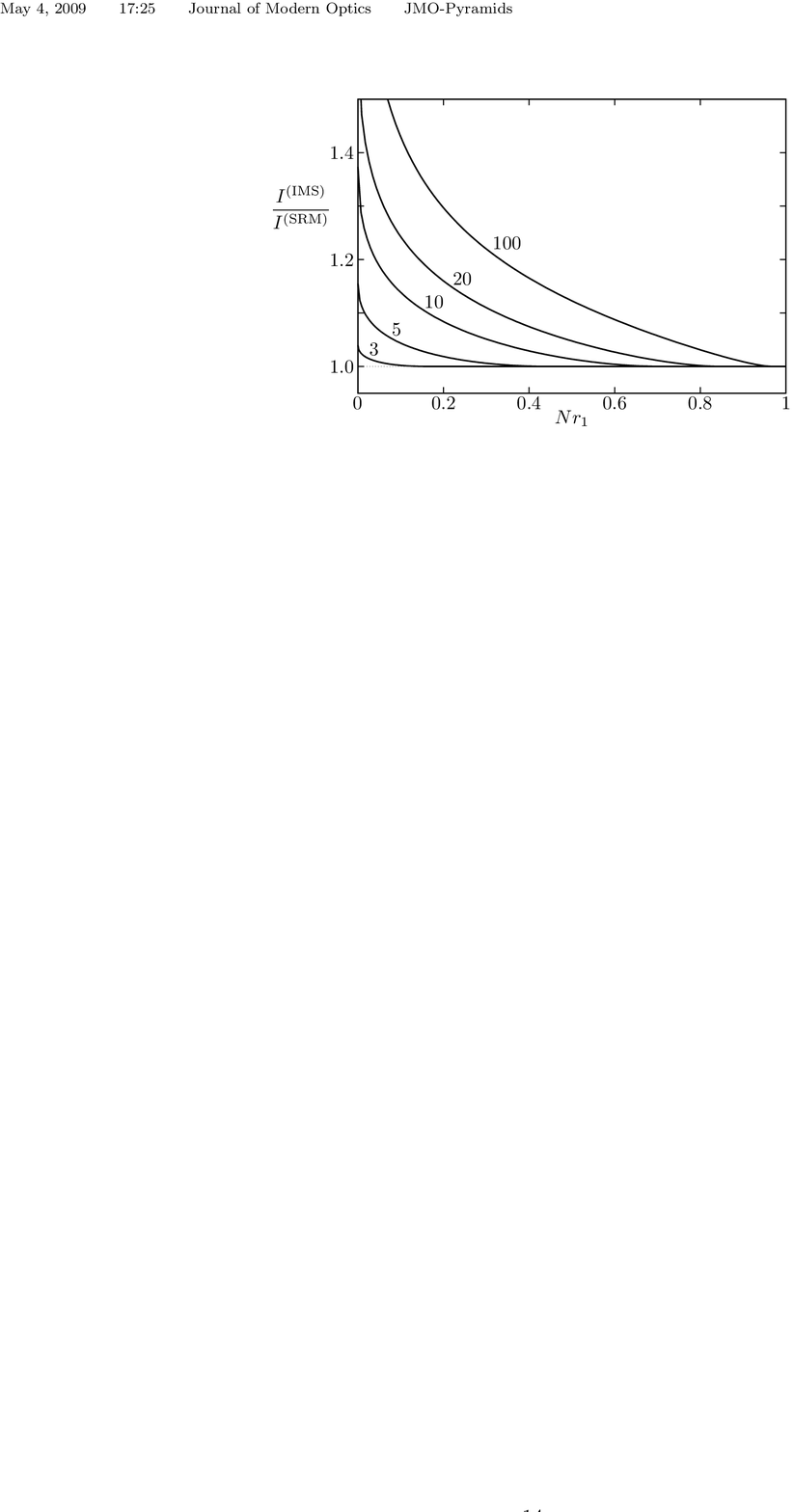}}
\caption{\label{fig:acute}%
Comparison of the accessible information and the information accessed by the
SRM. 
For $N=3$, $5$, $10$, $20$, and $100$ the lines show
$I^{\text{(IMS)}}$ in units of  $I^{\text{(SRM)}}$, for
acute pyramids, as a function of  $Nr_1$.
The figure is adopted from Ref.~\cite{englert}.
}
\end{figure}

Accordingly, the accessible information for acute pyramids is given by
\begin{equation}
  \label{acuteIMSinfo}
  I^{\text{(IMS)}}=\left\{
    \begin{array}{c@{\ \mbox{for}\ }l}
      I^{\text{(SRM)}}\mbox{\ of (\ref{SRMinfo})} & 
      \displaystyle r_0<\frac{4N-4}{N^2} \\[2.5ex]
      \displaystyle \frac{N-Nr_0}{N-2}\log_2(N-1) &
      \displaystyle r_0>\frac{4N-4}{N^2} 
    \end{array}\right.
\end{equation}
As demonstrated in Fig.~\ref{fig:acute} for pyramids with few 
as well as many edges,
the SRM coincides with the IMS for pyramids with a shape close to the
orthogonal pyramid (${Nr_1=1}$), but accesses only a small
fraction of the accessible information if the pyramid has many edges and a
sufficiently small volume.  
This parameter regime is relevant in the security analysis for some schemes
for quantum key distribution \cite{englert}.

\subsection{Obtuse pyramids}
\label{sec:obtuse}

\begin{figure}[b]
\centerline{\includegraphics[bb=190 595 360 740,clip=]{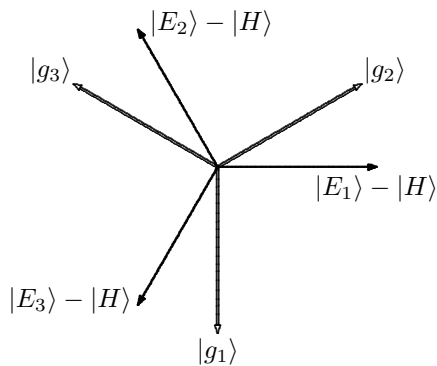}}
\caption{\label{fig:rotated}%
Shor's flat pyramid with three edges $\ket{g_j}$ is obtained from the
base pyramid with edges ${\ket{E_j}-\ket{H}}$ by a counter clockwise rotation
by $\pi/6$. 
All vectors lie in the plane perpendicular to the height ket $\ket{H}$. }
\end{figure}

The IMS for obtuse three-edge pyramids with very small volume was found by
Shor \cite{shor}.  
It consists of six rank-1 outcomes that make up two pyramids:
an almost flat lifted pyramid $|\bar{e}_j\rangle$, and
a flat pyramid with edges $|g_j\rangle$ that, in Shor's description, is
obtained by rotating the edges $|E_j\rangle-|H\rangle$ of the pyramid base 
by $30^\circ$ in the plane perpendicular to $|H\rangle$, 
see Fig.~\ref{fig:rotated}.
This rotation is, however, a particular feature of the ${N=3}$ case.
Another view that is more useful for going from ${N=3}$ to ${N=4,5,\dots}$
regards the kets $\ket{g_j}$ as the difference kets of the pyramid edge
kets, 
\begin{equation}
  \label{gShor}
  \ket{g_1}\propto\bigl(\ket{E_2}-\ket{H}\bigr)-\bigl(\ket{E_3}-\ket{H}\bigr)
           =\ket{E_2}-\ket{E_3}\propto\bket{[23]}
\end{equation}
and likewise $\ket{g_2}\propto\bket{[31]}$ and $\ket{g_3}\propto\bket{[12]}$.

This observation is the clue to finding the IMS for obtuse pyramids.
In close analogy to what worked for acute pyramids, we turn to the MUD for
obtuse pyramids and release the MUD condition $t=\sqrt{r_1/r_0}$.
Combining (\ref{2ndPOM}) with the decomposition (\ref{sum[mn]}), we have 
a POM with ${N+N(N-1)/2=N(N+1)/2}$ outcomes,
\begin{equation}
  \label{IMSobtuse}
   \Pi_k=\ket{\bar{e}_k}\frac{1}{t^2}\bra{\bar{e}_k}
   \quad\mbox{and}\quad
   \Pi_{[mn]}=\bket{[mn]}\frac{2}{N}\frac{t^2-1}{t^2}\bbra{[mn]} 
\end{equation}
where $k,m,n=1,2,\dots,N$ with the restriction $m<n$ to avoid double counting.
The optimal value of $t$ is determined numerically, and we find $t=1$ (SRM)
except when 
\begin{equation}
  \label{threshold}
  \begin{array}{l@{\quad\mbox{for}\quad}l}
    Nr_0<0.1837 & N=3\,,\\[0.5ex]
    Nr_0<0.0870 & N=4\,,\\[0.5ex]
    Nr_0<0.0286 & N=5\,,\\[0.5ex]
    Nr_0<0.0028 & N=6\,.
  \end{array}
\end{equation}
An extensive numerical search for the IMS with the aid of the iterative
steepest-ascent method of Ref.~\cite{rehacek} confirms that the IMS is indeed
of the form (\ref{IMSobtuse}).

\begin{figure}
\centerline{\includegraphics[bb=135 560 420 750,clip=]{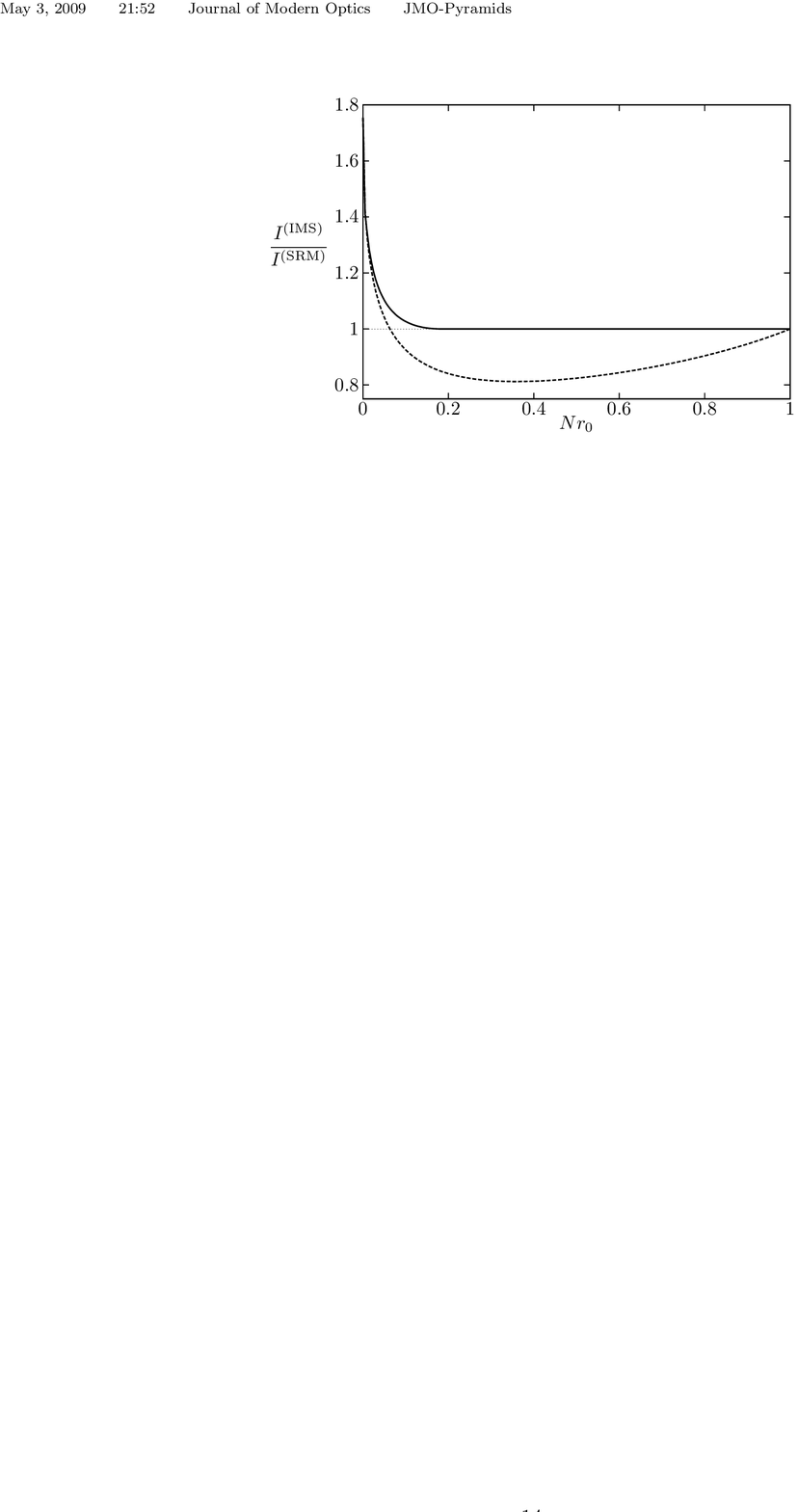}}
\caption{\label{fig:obtuse}%
Comparison of the accessible information and the information accessed by the
SRM, for obtuse pyramids with ${N=3}$ edges. 
The solid line shows $I^{\text{(IMS)}}$ in units of  $I^{\text{(SRM)}}$
as a function of $Nr_0$.
The broken line reproduces $I^{\text{(MUD)}}$ of the left plot in
Fig.~\ref{fig:MUDvsSRM}.  
}
\end{figure}

Figure~\ref{fig:obtuse} illustrates the matter for ${N=3}$,
where the information accessed by the MUD is also shown for comparison.
For $N=4$, $5$, or $6$, the respective plots are similar but the range of $Nr_0$
where $I^{\text{(IMS)}}$ exceeds $I^{\text{(SMR)}}$ is smaller, and the
improvement of the IMS over the SRM is not as substantial as for ${N=3}$.

While the MUD outperforms the SRM for $Nr_0\ll1$, it does not access as much
information as the IMS, except for the flat pyramids (${Nr_0=0}$), for which
the MUD and the IMS coincide.
The limiting value for $r_0\to0$ is thus given by (\ref{MUD/SRMlimit}), equal
to $1.755$ for ${N=3}$, as confirmed by Fig.~\ref{fig:obtuse}.

\section{Unified view}
\label{sec:general}

The various POMs discussed above are composed of a small set of rank-1
outcomes: 
the projector on the height ket $|H\rangle$ that is aligned with the axis of
the pyramid, the projectors on the $N$ edge kets  $|\bar{e}_k\rangle$ of a
lifted pyramid with its implicit dependence on parameter $t$, and the
projectors on the $N(N-1)/2$ difference kets $\bket{[mn]}$. 
A unified view regards all POMs as special cases of this decomposition of the
identity:
\begin{equation}
\label{allPOMs}
|H\rangle\frac{w_1}{r_0}\langle H|+
w_{2}\sum_k|\bar{e}_k\rangle\langle\bar{e}_k|
+\frac{2w_3}{N}\sum_{m<n} \bket{[mn]}\bbra{[mn]}=1\,,
\end{equation}
where the nonnegative weights $w_1$, $w_2$, and $w_3$  are subject to
\begin{equation}
\label{constraints}
w_1+t^2w_{2}=1\quad \text{and}\quad w_{2}+w_3=1\,,
\end{equation}
so that a particular POM is specified by stating the value of $t$ and the
value of one of the three weights.
Any permissible choice of $w_1$, $w_2$, $w_3$, and $t$ defines a POM with at
most $1+N(N+1)/2$ outcomes, one outcome for each summand in (\ref{allPOMs})
that carries a positive weight. 
The accessed information is 
\begin{equation}
\label{generalinfo}
\begin{split}
I=&w_{2}\biggl[
\frac{\bigl(t\sqrt{r_0}+(N-1)\sqrt{r_1}\,\bigr)^2}{N}
\log_2\frac{\bigl(t\sqrt{r_0}+(N-1)\sqrt{r_1}\,\bigr)^2}{t^2 r_0+(N-1)r_1}\\
&\phantom{w_{2}\biggl[}
+\frac{N-1}{N}\bigl(t\sqrt{r_0}-\sqrt{r_1}\,\bigr)^2
\log_2\frac{\bigl(t\sqrt{r_0}-\sqrt{r_1}\,\bigr)^2}{t^2 r_0+(N-1)r_1}
\biggr]\\
& +w_3(1-r_0)\log_2\frac{N}{2}\,,
\end{split}
\end{equation}
where we have no contribution from the inconclusive outcome with weight $w_1$. 

\begin{table}
\tbl{The various POMs as particular cases of (\ref{allPOMs}).}
{\begin{tabular}{@{}lp{200pt}@{}}\toprule
POM & {\raggedright parameter values}\\ \colrule    
SRM & {\raggedright $t=1$, $w_2=1$}\\
MUD & {\raggedright $t=\sqrt{r_1/r_0}$, $w_2=\min\{1,r_0/r_1\}$}\\
IMS & {\raggedright ${r_0>r_1}$: $t$ from (\ref{OPTacute}), $w_2=1$}\\
    & {\raggedright ${r_0<r_1}$: $w_2=1/t^2$ and $t=1$, except when $Nr_0$ is
      below the threshold value of (\ref{threshold}), in which case the value
      of $t>1$ is found by a simple numerical search}\\   
\botrule\end{tabular}}  
\label{tbl:summary}
\end{table}

As a summary, Table~\ref{tbl:summary} lists the parameter values for the SRM,
the MUD, and the IMS. 
With regard to the historical conjecture that the square-root measurement
extracts the accessible information, we conclude that this is true for
pyramids with large volume, but not for small-volume pyramids that either are
acute and have more than two edges or are obtuse and have three, four, five,
or six edges.

\section*{Acknowledgments}
We are very grateful for the valuable discussions with 
Dagomir Kaszlikowski, Ajay Gopinathan, Frederick Willeboordse, Shiang
Yong Looi, and Sergei Kulik.
J.~\v{R} wishes to thank for the kind hospitality
received during his visits to Singapore.
This work was supported by Grant MSM6198959213 of the Czech Ministry of 
Education, and by NUS Grant WBS: R-144-000-109-112.
Centre for Quantum Technologies is a Research Centre of Excellence funded by
Ministry of Education and National Research Foundation of Singapore.

\newcommand{\arXiv}[2][quant-ph]{\textrm{eprint\ arXiv:#1/#2}}
\newcommand{\arxiv}[2][quant-ph]{\textrm{eprint\ arXiv:#2 [#1]}}


\begin{thebibliography}{00}
\bibitem{range}
We regard the identity operator on the right-hand side of (\ref{genPOM}) as
the projector on the joint range of the signal states $\rho_j$ and ignore the
orthogonal complement of the Hilbert space of kets, if there is one.
\bibitem{isometry}
We note that, if the underlying Hilbert space is infinite dimensional, 
$U$ must be isometric but not necessarily unitary.
\end{thebibliography}
\end{document}